\documentclass[twocolumn,amssymb]{revtex4}

\usepackage{graphicx}
\usepackage{bm}

\begin{document}


\title{Measurement of the magneto-optic properties of thin-film metals and high temperature superconductors}



\author{J. \v Cerne}
\affiliation{Department of Physics, University of Maryland, College Park, MD 20742, USA}
\affiliation{Department of Physics, University at Buffalo, The State University of New York, Buffalo, NY 14260, USA}

\author{D.C. Schmadel}
\affiliation{Department of Physics, University of Maryland, College Park, MD 20742, USA}

\author{L. Rigal}
\affiliation{Laboratoire National des Champs Magn\'{e}tiques Puls\'{e}s, Toulouse, France}
\affiliation{Department of Physics, University of Maryland, College Park, MD 20742, USA}

\author{H.D. Drew}
\affiliation{Department of Physics, University of Maryland, College Park, MD 20742, USA}
\affiliation{Center of Superconductivity Research, University of Maryland, College Park, Maryland 20742}


\date{November 13, 2002}

\begin{abstract}
A sensitive polarization modulation technique uses photoelastic modulation and hetrodyne detection to simultaneously measure the Faraday rotation and induced ellipticity in light transmitted by semiconducting and metallic samples.  The frequencies measured are in the mid-infrared and correspond to the spectral lines of a CO$_2$ laser.  The measured temperature range is continuous and extends from 35 to 330K.  Measured samples include GaAs and Si substrates, gold and copper films, and YBCO and BSCCO high temperature superconductors.
\end{abstract}



\maketitle


\section{Introduction}

Conventional dc Hall effect measurements in novel electronic materials such as high temperature superconducting cuprates (HTSC) have been essential in revealing the unusual character of these systems.~\cite{dchall} Extending these measurements to higher frequencies~\cite{achall} allows one to probe more effectively the energy scales of the
system (e.g., the plasma frequency, the cyclotron frequency and the carrier relaxation rates) while minimizing the effects of impurity scattering, which can dominate dc measurements. Further, the Hall angle and Hall conductivity satisfy sum rules~\cite{DrewColemanSumRules}, which when applied to ac Hall data provide insight into the intrinsic electronic structure of systems ranging from conventional Fermi liquid metals~\cite{landau} to more exotic metals such as HTSC~\cite{htsc} and other transition metal oxides.~\cite{cmr} 

Magneto-polarimetry measurements can be used to extend Hall effect measurements into the infrared frequency range ($10^{13}$~Hz).  These measurements are sensitive to the complex Faraday angle $\theta_{\text{F}}$, which is closely related to the complex Hall angle
$\theta_{\text{H}}$.~\cite{au}  Though $\theta_{\text{F}}$ and $\theta_{\text{H}}$ tend to be small for metals in the mid-infrared
(MIR, 900-1100 cm$^{-1}$), there
are a number of advantages in performing these higher frequency
measurements. First, the high frequency  allows
one to avoid impurity scattering or grain boundary effects which may
dominate lower frequency Hall measurements. This is especially important in new
materials which often contain many impurities and defects. Furthermore, the MIR measurements allow one to examine the trends
observed at lower frequencies. Since $\tan\theta_{\text{H}}$ (and $\theta_{\text{H}}$) 
obeys a sum rule~\cite{drew} it is
very useful to be able to integrate $\theta_{\text{H}}$ to higher frequencies to
verify whether (and where) the Hall angle sum rule saturates or whether
there is more relevant physics at even 
higher frequencies.  Finally, since the high
frequency behavior of $\theta_{\text{H}}$ is constrained by the general
requirements of response functions, a simple, model-independent asymptotic form for $\theta_{\text{H}}$ becomes more accurate at 
higher frequencies.

In this paper, we present a sensitive mid-infrared polarimetry technique, which uses photo-elastic modulation and heterodyne detection, and which can be used to explore a variety of materials.~\cite{cerne,schmadel} In what follows we shall first introduce the hardware components and describe their operation, next we shall present an analysis of both the operation and data, and finally we shall present recent results on semiconductors, metal films, and high temperature superconductors.

\section{Experimental System}

\subsection{Overview}

\begin{figure*}
\includegraphics{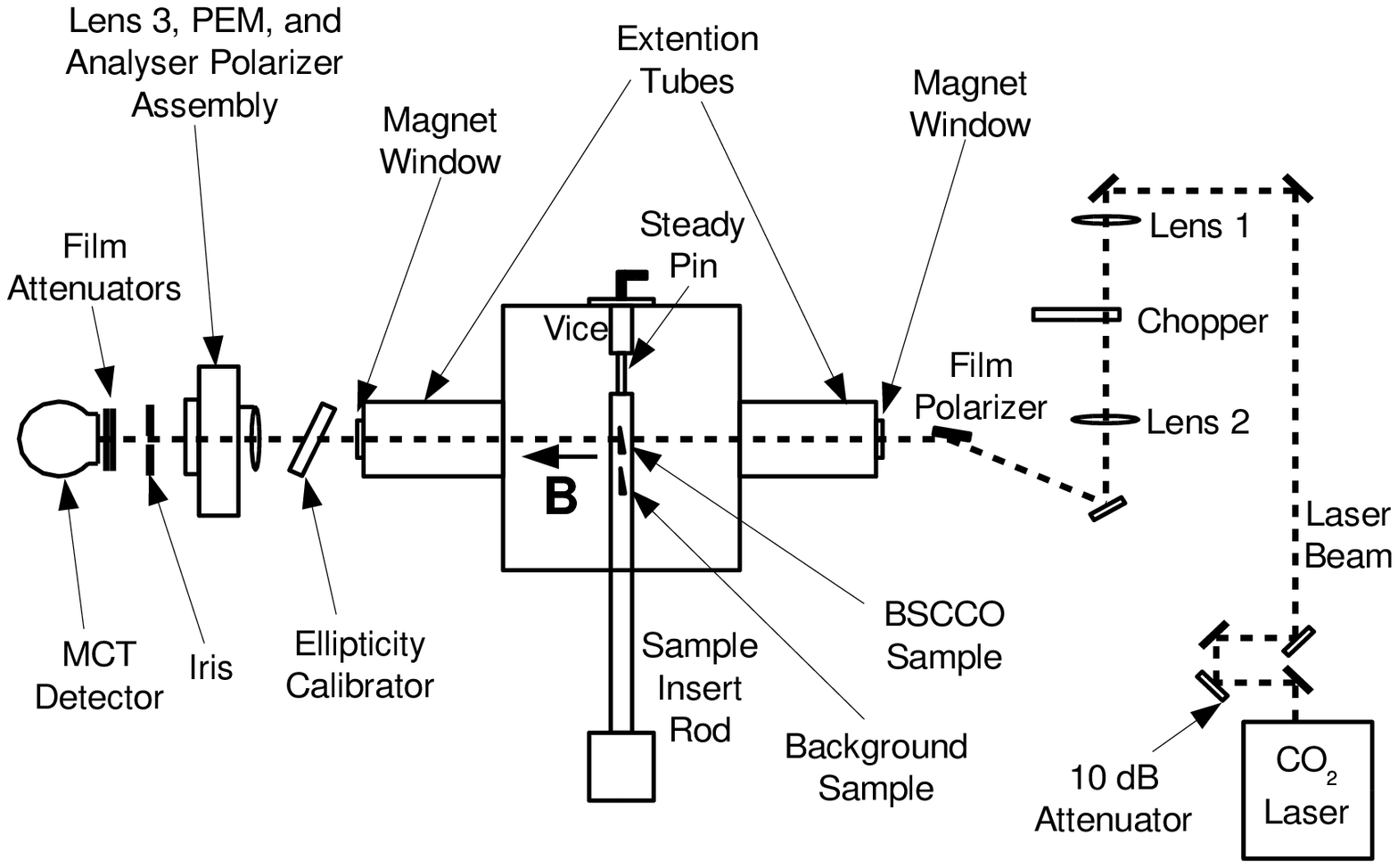}
\caption{\label{fig;opticalPath} Overall schematic of the optical path.  The the dotted line represents the CO$_2$laser beam.  The CO$_2$ laser and the first four optic elements are actually located on a separate mount detached from the optical table.}
\end{figure*}

\begin{figure}
\includegraphics{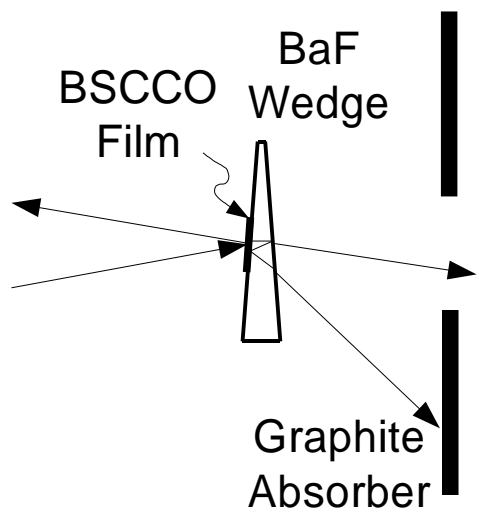}
\caption{\label{fig;sample} Wedged sample and beam path.  The paths of the reflected beams are sufficiently separated by the 2$^\circ$ wedge to prevent their entering the detection system or reentering the laser.}
\end{figure}

The experimental system of the current work measures the very small real and imaginary parts of the complex Faraday angle imparted to CO$_2$ laser radiation traveling perpendicular to and transmitted by the sample film which is placed in a perpendicular magnetic field.  During operation the experimental system performs four major tasks essentially simultaneously:
\newcounter{task}
\begin{list}{\arabic{task})}
{\usecounter{task}}
\setlength{\leftmargin}{0.7cm}
\setlength{\rightmargin}{0.5cm}
\item generating and directing a monochromatic, linearly polarized light beam normal to the sample
\item producing and controlling a magnetic field at the sample normal to its surface
\item setting and maintaining the temperature of the sample
\item analyzing the portion of the light beam transmitted by the sample to determine the complex Faraday angle.
\end{list}
Figure~\ref{fig;opticalPath} schematically illustrates the optical path in which the light beam initially issues linearly polarized from the CO$_2$ laser and, after enduring various steering and attenuation components, proceeds through lens 1, which focuses it to a point at the chopper.  The chopper impresses onto the intensity of the beam a temporal square wave of ~112 Hertz which will later facilitate the removal of laser power variations using ratios.  Lens 2 then refocuses the emerging beam so as to eventually produce a focused diffraction spot on the surface of the sample.  The intervening element, a film polarizer, ``cleans up'' the beam's polarization, removing any contamination caused by the chopper, attenuators, etc.  It is this highly linearly polarized beam, which after passing through the ZnSe magnet housing window encounters the sample as shown in Fig.~\ref{fig;sample}. The sample will reflect, absorb, and transmit portions of this beam.  The portion transmitted will have sustained a Faraday rotation proportional to the magnetic field and consistent with the physics peculiar to the sample. We shall assign the x axis to the initial polarization direction and the z axis to the direction of propagation.

The Faraday rotation angle includes both real and imaginary terms.  The real term corresponds to a simple geometric rotation of the polarization vector about the direction of propagation.  The imaginary term relates directly to the ellipticity of the polarization and has been called the circular dichroism.  The magnitude of the Faraday rotation at CO$_2$ wavelengths is minute---about \(10^{-4}\) radians.  As such, one may consider the electric vector along the y axis to be a feeble signal added to the comparatively prodigious signal of the original the electric vector along the x axis.   This small signal cannot be measured directly  by the common method of cross polarizers because the power in the field polarized along the y axis is proportional to the square of the Faraday angle or ~\(10^{-8}\) times the power in the original beam in the x polarization.  Clearly, leakage of the input beam through either polarizer would effectively direct power into the y axis polarization which would well overwhelm any amount occasioned by the sample.  The natural choice in such instances is heterodyne detection.

Heterodyne detection produces cross terms which are proportional to the electric vector along the y axis rather than its square.  We realize heterodyne detection by phase modulating the electric vector along the y axis using a photoelastic modulator(PEM).  The PEM modulator is a model 90 manufactured by Hinds Instruments, Inc.  with a modulation frequency of 50 kHz.  It comprises essentially a block of ZnSe and a piezoelectric transducer which creates the stress alternations at a frequency of 50 kHz and ultimately the phase modulation of the light beam.  The electric vector along the x axis is essentially unaffected by the PEM.

Returning to our discussion of the optical path the beam emerges from the PEM and then strikes a polarizer on a 45$^\circ$ angle, which, simply stated, allows a portion of the electric vector from each of the y and x directions to contribute to that which emerges now at a 45$^\circ$ angle.  These two contributions after being mixed by the detector produce the 50 kHz PEM frequency along with sidebands.  The  amplitude of each of the sidebands is proportional to the electric vector along the y axis and also proportional to the corresponding Bessel function whose argument is the depth of modulation.  In fact, as will be derived below, the even sidebands or harmonics are proportional to the real part of the Faraday angle and the odd sidebands or harmonics are proportional to the imaginary part of the Faraday angle.

In the following sections we consider certain elements or subassemblies of the experimental system in detail beginning with the CO$_2$ laser.

\subsection{CO$_2$ Laser}

The CO$_2$ laser, fashioned from a model PL5 manufactured by Edinburgh, provides a number of spectral lines ranging from 9.174 to 10.860 microns (920 to 1090 cm$^{-1}$), any one of which is selected by a grating internal to the cavity.  A separate CO$_2$ laser spectrometer verifies the wavelength of each line.  The direction of the beam exiting the laser, however, differs for different lines sometimes by nearly as much as one half of a spatial mode.  Without realignment the result would be a spatial shift of the diffraction spot at the sample by as much as one half of a spot diameter.   The small size of a sample can aggravate the sensitivity of the system to such slight misalignment particularly when such changes cause the beam to wander off or even near the edge of the sample.  Lens 3 essentially focuses an image of the sample onto the iris shown in Fig.~\ref{fig;opticalPath} and may be used as and aid in regaining alignment as is necessary to compile valid relative frequency data.  The design and construction of the hardware well ensures that the relative position of the sample, lens 3, and the iris are constant.  Therefore, adjusting the laser beam steering to maximize the power through the iris after each laser line change likewise ensures that the beam is passing through the same area of the sample.  While actually taking data the iris is dilated to accommodate a small amount of beam wander.

Another consideration of the laser is the high output power---2 to 30 Watts.  Three attenuators reduce the power $\sim$30dB to avoid heating the sample while still producing a robust signal within the linear range of the MCT detector.  These attenuators are ZnSe windows having an antireflection coating on one side.  They  replace mirrors in the beam steering assemblies with the uncoated side serving as the reflective surface.  Each of these attenuators introduces approximately 10 dB of loss.   

The output power of the laser, notwithstanding its hardiness,  fluctuates temporally.  Therefore, producing usable data requires the formation of simultaneous ratios between the 2nd and 3rd harmonics, and a sample of the laser power level.  The current experiment employs the usual method of "source compensation"~\cite{ratio} which involves the chopper along with an additional lockin amplifier.  Note also that the heterodyne detection system described above permits use of the same detector for the power level sample as well as the 2nd and 3rd harmonics.  As to be derived below, the power level sample, aside from the chopper frequency, corresponds to the original optical carrier frequency centered within the sidebands mentioned above.  This feature eliminates the wavelength dependence of the detector and thus facilitates accurate relative measurements of the wavelength dependence of the complex Faraday angle.  A final consideration regarding the laser involves its placement with respect to the magnet.  The particular position of the laser is perpendicular to the magnetic field and at such a distance as to reduce detuning of the laser cavity caused by magnetostriction to a tolerable amount.

\subsection{Magnet System and Sample Mounting}

\begin{figure*}
\includegraphics{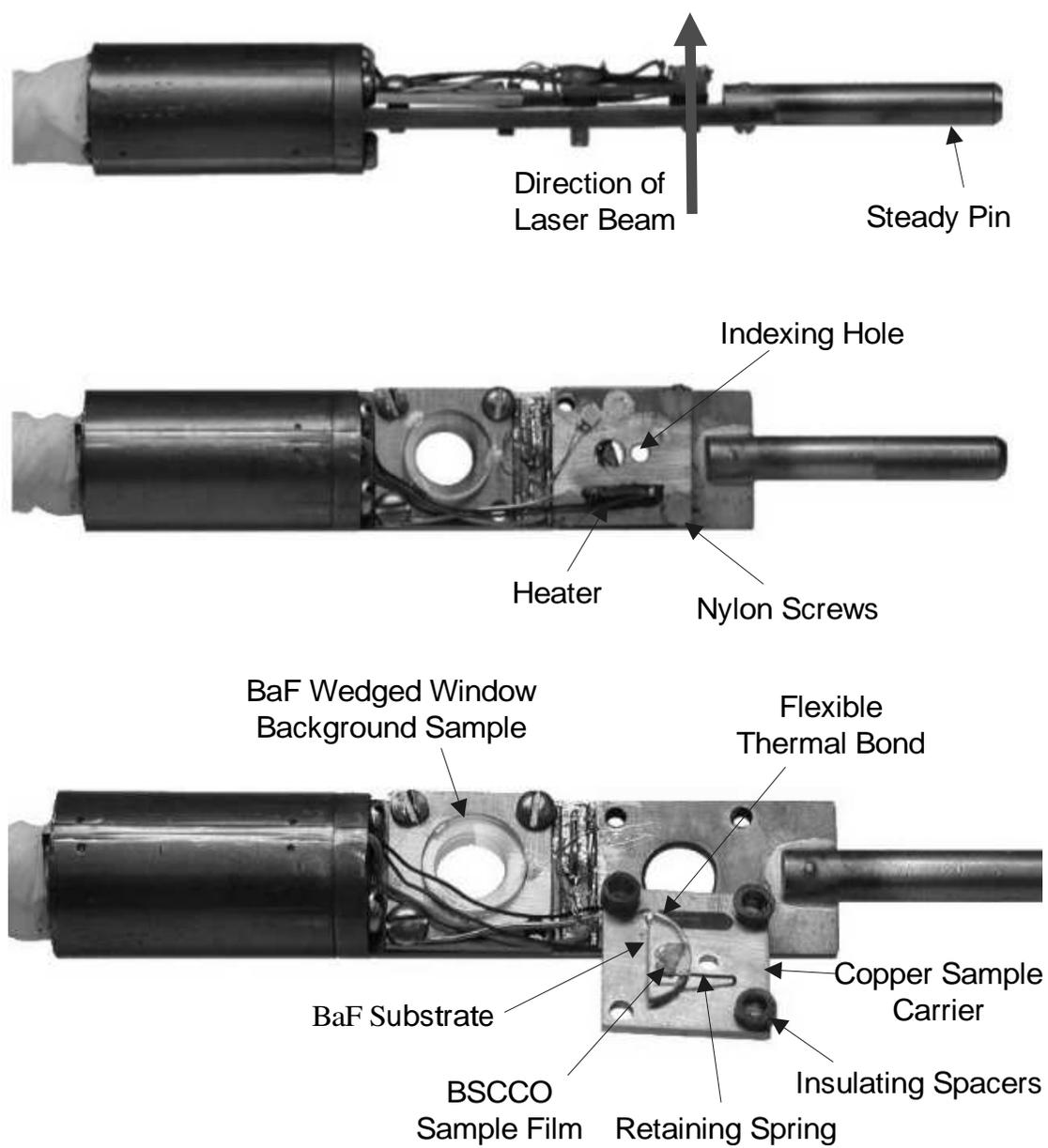}
\caption{\label{fig;sampleStick} Sample mount assembly}
\end{figure*}

The magnet system began as an 8 Tesla, split coil, Helmholtz Spectromag manufactured by Oxford Instruments and, except for modifications to the external windows, the sample handling hardware, and the internal bore tube shielding, it has served the experiment with little modification.  To understand the modifications to the external windows, consider that the wavelength range of interest requires ZnSe windows.  However, the band gap of ZnSe is only 2.7eV, and the magnetic field of 2 Tesla at the original location of the windows caused a significant Faraday rotation, which overwhelmed that of the sample.  The extension tubes locate the windows where the magnetic field is less than 0.1 Tesla.  The remaining Faraday rotation background is thus reduced to the same order as that of the sample and having been carefully measured for different wavelengths can be easily removed from the data at a later time.  For magnets requiring cold windows 
located within the high field region the background may be reduced by using a material with a higher band gap such as BaF. 
The sample handling hardware, ill-suited to the task at hand, enjoyed substantial redesign and remanufacture. The effort included adding a steady pin and vise, implementing a more responsive temperature control system, and developing a near stress free sample mount.  The steady pin, depicted in Fig.~\ref{fig;sampleStick} protrudes from the blade of the original sample stick to which it is brazed.  The vise, located within the magnet housing, engages the pin using a teflon collet.  This combined apparatus restrains the sample stick against the forces induced by the magnetic field and thus prevents the large, extraneous interference signals resulting from changing multi-paths and etalons.

Figure~\ref{fig;sampleStick} also depicts parts of the new temperature control system which comprises a heater and a cooling link.  The heater is a 400 Ohm 1/4 Watt metal film resistor potted into the copper sample carrier using 2850FT Epoxy with Catalyst 9 both of which are manufactured by Emerson \& Cuming.  The cooling link is simply a 4 cm length of 22 gauge copper wire connecting the copper sample carrier to the blade of the original sample stick from which it is otherwise thermally insulated by spacers and nylon screws.  The operation is delightfully simple:  Liquid helium delivered to the sample stick by the original provision, cools the blade to about 10 K, and the cooling link cools the copper sample carrier.  Current delivered to the resistor in an easily controlled fashion can provide up to 2 Watts of heat to the copper sample carrier.  Because of the low thermal mass of both the copper sample carrier and the link, a compromise temperature emerges within about one minute.  Sweeping the temperature entails nothing more than adjusting the current.  A important advantage of this system is that only the rather small copper sample carrier changes temperature.  The remaining hardware, remaining essentially constant in temperature, finds little reason to move or warp so as to adversely affect the measurement.

The near stress-free sample mount consists of a phosphor bronze wire retaining spring and a thermally conductive flexible silver-filled RTV known as Eccobond 59C manufactured by Emerson \& Cuming, Inc.  The particular sample shown attached to the sample stick is a small irregularly shaped film of optimally-doped Bi$_2$Sr$_2$Ca$_1$Cu$_2$O$_{8+\delta}$ approximately 200 nanometers thick which had been peeled from a single bulk crystal.  This small film was placed against a polished surface of a barium fluoride crystal which serves as a substrate and maintains the temperature of the sample.  Van der Waal's force holds the film in place.  A 2$^\circ$ wedge of the BaF substrate eliminates etalon effects.  Only one of the two corners of the substrate is cemented to the copper sample carrier.  The phosphor bronze retaining spring lightly holds the other corner while allowing some motion to relieve the stress caused by dissimilar thermal expansion coefficients.  Without this provision the stress induced in the substrate had caused overwhelming and unpredictable complex Faraday rotations.  It is this stress free mounting which actually facilitated the fast temperature scans required to eliminate 1/f noise apparent in previous work.  The copper sample carrier is provided with a small indexing hole also shown in Fig.~\ref{fig;sampleStick}.  Prior to insertion of the sample stick into the magnet the sample's position is accurately measured with respect to this hole.  After insertion, the indexing hole is located using the transmitted intensity of the laser beam.  The sample is then positioned within the beam by raising or lowering the magnet and sliding the sample stick in or out.  Finally the new reentrant bore tubes are fit with graphite plates to absorb the stray radiation scattered from the incident laser beam by the various reflective surfaces, e.g., the sample, the substrate, the magnet windows, etc.

\subsection{Optical Table Components }

The remaining items of Fig.~\ref{fig;opticalPath} which along with the magnet are located on the aluminum optical table will receive consideration within this section.  Among these items are the chopper and lens 1.  As explained above, the chopper impresses upon the beam a square wave amplitude variation of 112 Hertz.  Lens 1 participates in this task by focusing the beam to a point at the chopper blades.  Thus, the passage of a blade cuts the beam off and on in an most abrupt manner.  This prevents spatial variations within the beam from causing phase and amplitude errors in the reported laser power level. 

Also of particular importance a pivot platform upon which sits the optical detection system.  Two aluminum box beams connect this platform to a vertical pivot located directly below the sample.  When the two clamps securing the platform to the optical table are loosened the platform may be rotated about the vertical axis of the pivot.  This motion is necessary to align the detection system to both the indexing hole and wedged substrates which bend the laser beam about the same axis by an amount depending on the substrate wedge angle and its index of refraction.

\subsection{Optical Detection System }

The first component encountered by the laser beam after passing through the magnet window is lens 3 which focuses an image of the sample onto the iris as previously discussed.  This lens joins with the PEM and polarization analyzer in an assembly, which can be rotated as a single unit about the input optical axis, which is the z axis.  With a moment's thought one will recognize that such a rotation is equivalent to a real, but opposite, Faraday rotation, at least for sufficiently small angles so that the polarization sensitivity of the MCT detector is not apparent.  Fixed calibration rotation stops limit this rotation to a known amount thus serving as a Faraday rotation calibrator.  Simply put, one rotates the assembly the known amount and uses this to scale the empirical values for each wavelength.

Another important consideration involves reflection at the surfaces of the ZnSe interaction block of the PEM.  A reflected beam which makes additional passes through the ZnSe interaction block will receive additional modulation.  Since, the cross term is a function of the depth of modulation, such triply modulated stray beams can cause significant errors.  An AR coating and a tilting of the PEM by 25$^{\circ}$ reduce and displace reflected beams and, thereby, sufficiently reduce their effect.

A variable selection of polyimide films further attenuate the laser beam so that its power is within the linear range of the MCT detector which is a model J15D14 mercury cadmium telluride detector manufactured by EG\&G Judson.

\subsection{Electronic Instrumentation}

Along with the electronics which attend the magnet, PEM, chopper, etc., the system uses three model 7260 harmonic lockin amplifier/detectors manufactured by EG\&G.  These lockins differ from earlier styles in that they detect and report the RMS voltage of a selectable harmonic of the input signal.  This feature is essential because, as mentioned, the even harmonics of the phase modulation of the PEM are proportional to the real part of the Faraday angle and the odd harmonics are proportional to the imaginary part.  The first lockin determines the RMS voltage at 112 Hertz and is usually considered to be the DC reference.  The second lockin determines the RMS voltage at 2$\omega$, the 2nd harmonic of the PEM frequency, and the third determines the RMS voltage at 3$\omega$, the 3rd harmonic.  A Labview program running on a local computer records data points consisting of the following:
\begin{eqnarray*}
\left\{\text{time\ or}\ {\bf B} \text{\ field},\ \frac{V_{\text{2}\omega}(x)}{V_{\text{112}}},\,\frac{V_{\text{2}\omega}(y)}{V_{\text{112}}},\ \right.\\\left.\frac{V_{\text{3}\omega}(x)}{V_{\text{112}}},\,\frac{V_{\text{3}\omega}(y)}{V_{\text{112}}},\, V_{\text{112}}\right\}.
\end{eqnarray*}
If the beam moves within the PEM aperture the phase of the harmonics will change.  This phase is initially set such that the $y$ channel is very nearly zero with most of the signal in the $x$ channel.  To avoid any errors from the changing phase we use the positive definite magnitude formed from the combination of the $x$ and $y$ channels and apply the sign chosen determined by the calibration procedure described later. 

\section{Analysis}

In the analysis of both the system and data we employ two formalisms.  One concerns the representation and transformation of the polarization state of light and the other concerns the transmission and reflection response of multilayer stacks with complex material properties, $\sigma$, $\epsilon$, or $\mu$. Both are presented as appendices.

\subsection{Faraday Angle $\theta_{\text{F}}$ versus Lockin Outputs}

For light initially propagating in the z direction and polarized along the x axis, we define the Faraday angle as 
\begin{equation}
\theta_{\text{F}}=\arctan \frac{t_{yx}}{t_{xx}}\approx \frac{t_{yx}}{t_{xx}}
\label{eq;FaradayDef}
\end{equation}
where $t_{xx}$ is the field transmission along the x polarization,
$t_{yx}$ is the field transmission along the y polarization, and the arctan function was dropped because of the very small angles
involved in the current work. As such
it represents the complex amplitude (amplitude and phase) in the y polarization having
been derived from the incident radiation in the x channel.  One will recognize $\text{Re} \left( \theta_{\text{F}}\right)$ as the geometric rotation of the incoming polarization and $\text{Im} \left( \theta_{\text{F}}\right)$ as the circular dichroism.

Consider next that the experimental system examines the light transmitted
by the sample, which results from incident light initially polarized
along the x axis (recall the z axis is the direction of propagation).
A simple matrix equation represents this activity in the linear polarization
basis as
\begin{equation}
\left( \begin{array}{cc}
t_{xx} & t_{xy}\\
t_{yx} & t_{yy}
\end{array}\right) \left( \begin{array}{c}
x_{\text{in}}\\
0
\end{array}\right) =\left( \begin{array}{c}
x_{\text{out}}\\
y_{\text{out}}
\end{array}\right) 
\label{eq;sampleTransmission}
\end{equation}
Assuming the a and b axes of the sample to be indistinguishable, i.e.,
near square symmetry, and noting that the B field is uniform and parallel
to the z axis, we know that the sample transmission of Eq.~(\ref{eq;sampleTransmission}) will be diagonal in the circular basis. Using the polarization formalism in the appendix the sample can be represented as
\begin{equation}
\left( \begin{array}{cc}
t_{p} & 0\\
0 & t_{n}
\end{array}\right) =\left( \begin{array}{cc}
t_{xx}-i\, t_{yx} & 0\\
0 & t_{xx}+i\, t_{yx}
\end{array}\right) 
\label{eq;sampleTransmissionCircular}
\end{equation}
where $t_{p}$ is the transmission coefficient corresponding to positive
helicity (positive $\widehat{\phi}$ rotation seen at fixed point for
a wave traveling in the positive z direction, and $t_{n}$ is the transmission
coefficient corresponding to negative helicity (negative $\widehat{\phi}$
rotation seen at fixed point for a wave also traveling in the positive
z direction. Transforming back to the linear basis the sample matrix becomes
\[\left( \begin{array}{cc}
t_{xx} & t_{xy}\\
t_{yx} & t_{yy}
\end{array}\right) =\left( \begin{array}{cc}
t_{xx} & -t_{yx}\\
t_{yx} & t_{xx}
\end{array}\right)
\nonumber
\]
which is considerably simpler.  It is also useful to note that by the definition in Eq.~(\ref{eq;FaradayDef}) the Faraday angle relates very simply to the ratio of the transmissions of the right and left circular polarization:
\[\left( \begin{array}{cc}
\frac{t_{p}}{t_{n}} & 0\\
0 & \frac{t_{n}}{t_{p}}
\end{array}\right) =\left( \begin{array}{cc}
e^{-i\, 2\, \theta _{\text{F}}} & 0\\
0 & e^{i\, 2\, \theta _{\text{F}}}
\end{array}\right).
\]
Next, for the purpose of developing a relation between the sample transmission
and the lockin outputs consider the relevant experimental elements
represented schematically as:
\begin{eqnarray*}
\text{laser} \rightarrow \text{sample} \rightarrow \text{PEM} \rightarrow \text{polarizer at}\  45^\circ \\
\rightarrow \text{square law detector }\rightarrow \ \text{lockins}.\ \ \ \ \ \ \ \ \ \ 
\end{eqnarray*}
Again using the polarization formalism, the components in the the first line, which operate on the laser beam $|x\rangle$ may be represented as:
\begin{eqnarray*}
{\bf R}\left( \frac{\pi }{4}\right) \times \left( \begin{array}{cc}
1 & 0\\
0 & 0
\end{array}\right) \times {\bf R}^{-1}\left( \frac{\pi }{4}\right) \times \text{PEM}\times  \text{sample}\times | x \rangle
\end{eqnarray*}
or
\begin{eqnarray*}
\left(\begin{array}{c}
\langle p|\text{out} \rangle\\
\langle n|\text{out} \rangle\\ 
\end{array}\right) =
\left( \begin{array}{cc}
e^{-i\frac{\pi }{4}} & 0\\
0 & e^{i\frac{\pi }{4}}
\end{array}\right) 
\left( \begin{array}{cc}
1 & 0\\
0 & 0
\end{array}\right) 
\left( \begin{array}{cc}
e^{i\frac{\pi }{4}} & 0\\
0 & e^{-i\frac{\pi }{4}}
\end{array}\right)\\\times 
\left( \begin{array}{cc}
1 & -i\\
1 & i
\end{array}\right) 
\left( \begin{array}{cc}
e^{i\, \beta \cos \omega t} & 0\\
0 & 1
\end{array}\right) 
\left( \begin{array}{cc}
1 & 1\\
i & -i
\end{array}\right)\\\times
\left( \begin{array}{cc}
e^{-i\theta _{\text{F}}} & 0\\
0 & e^{i\theta _{\text{F}}}
\end{array}\right) \left( \begin{array}{cc}
1 & -i\\
1 & i
\end{array}\right) \left( \begin{array}{c}
1\\
0
\end{array}\right) 
\end{eqnarray*}
where \( \omega = 2 \pi f\) is the radial frequency of the PEM modulation and $\beta$ is the PEM modulation depth or amplitude.  The above signal is incident upon a square-law detector whose output is a voltage proportional to the square of the modulus of the amplitude of the fields:
\begin{eqnarray*}
\text{voltage} \propto \langle \text{out}|\text{out} \rangle \propto \text{optical\ power}.
\end{eqnarray*}
For the small angles considered the multiplication of these matrices along with the Bessel function expansion of $e^{i\beta\cos \left( 2 \pi f \right) }$ produces
\begin{eqnarray}
\text{Re}\left(\theta_{\text{F}} \right) = \mathrm{Re}\left(\frac{t_{yx}}{t_{xx}}\right) = -\frac{1}{4 J_2(\beta)} \frac{V_{2 \omega}}{V_{\text{dc}}}
\label{eq;thetafVersusPower}
\\
\text{Im}\left(\theta_{\text{F}} \right) = \mathrm{Im}\left(\frac{t_{yx}}{t_{xx}}\right) = -\frac{1}{4 J_3(\beta)} \frac{V_{3 \omega}}{V_{\text{dc}}}
\label{eq;thetadVersusPower}
\end{eqnarray}
where $J_2(\beta)$ and $J_3(\beta)$ refer to 2nd and 3rd order Bessel functions respectively, $V_{DC}$ refers to the optical power at dc (actually at 112 Hz due to the chopper).

\subsection{Sample Properties versus Complex Faraday Angle}

Since the sample transmission matrix, Eq.~(\ref{eq;sampleTransmissionCircular})  is diagonal in circular polarization, each circular polarization channel acts independently and may be treated as such.  Indeed, by the same symmetry considerations, the sample matrix representation of the conductivity tensor is also diagonal in the circular basis.  In what follows we will develop an expression for the transmission of a sample-substrate combination and then separately apply this expression to each polarization channel.  This will give us an expression for the indices of refraction of the film for positive and negative helicity $n_{f,p}$ and $n_{f,n}$ respectively, which are then easily related to $\sigma_{xx}$ and $\sigma_{xy}$.

Figure~\ref{fig;sample} appearing earlier depicts the transmission through the sample, which is usually a film-substrate combination.  The input beam strikes the sample film, a portion reflects as shown, and a portion propagates into the sample film and after enduring some absorption arrives in the forward direction at the output side of the film.  This portion strikes the interface between the film and the substrate mounting surface.  Some reflects back and forth within the sample film and some proceeds as a beam into the wedged substrate soon reaching the substrate-air interface.   Most of this beam propagates into the air eventually arriving at the detection system.  Because of the wedge angle of the substrate-air interface, the small reflected portion leaves the optical path and is absorbed by strategically placed graphite slabs located in the reentrant bore tubes.  For this reason the substrate-air interface does not participate in the Faraday rotation for any of the materials of current interest and need not be considered.  All of the relevant activity above can therefore be represented schematically as
\begin{eqnarray*}
\begin{array}{c}
\text{air}\\
\longrightarrow 
\end{array}|\begin{array}{c}
\text{film}\\
\longleftrightarrow 
\end{array}|\begin{array}{c}
\text{substrate}\\
\longrightarrow 
\end{array}
\end{eqnarray*}
The formalism for multilayer transmission presented in the appendix represents this sequence generally as
\begin{eqnarray}
\left( \text{sample}\right) \left( \begin{array}{c}
E_{\text{air},\, \rightarrow }\\
E_{\text{air},\, \leftarrow }
\end{array}\right) =\left( \begin{array}{c}
E_{\text{sub},\, \rightarrow }\\
0
\end{array}\right) .
\label{eq;sampleMultilayer}
\end{eqnarray}
where $(\text{sample})={\bf S}_{\text{film}, \text{sub}}={\bf U}_{\text{film}}{\bf S}_{\text{air}, \text{film}}$ with {\bf S} denoting the interface matrix and {\bf U} denoting the propagation matrix.  That is, the sample acts on the electric field vectors  $E_{\text{air},  \rightarrow}$ and $E_{\text{air},  \leftarrow}$ of the incoming and reflected waves respectively on the input side producing the electric field vectors $E_{\text{sub},  \rightarrow}$ and $E_{\text{sub},  \leftarrow}$ of the outgoing and reflected waves on the output side, where, of course, we have no incoming wave. We desire only the transmission which from Eq.~(\ref{eq;sampleMultilayer}) is the (1, 1) element of the inverse of the sample matrix.  Since the transmission is the same for either propagation direction; a moments thought will lead one to propagate backwards to avoid taking the inverse of the sample matrix.  The end result is the transmission is given by the inverse of the (2, 2) element of the sample matrix, which using the relations in the appendix can be related (after some algebraic manipulation) to the material properties:
\begin{eqnarray}
&&t(n)=4n_{s}n_{f}e^{ikd}\nonumber\\
&&\times \left[ n_{s}\left( e^{i2kd}\left( n_{f}-1 \right) +n_{f}+1 \right)\nonumber\right.\\
&&\left.-n_{f} \left( e^{i2kd} \left( n_{f}-1 \right) -n_{f}-1 \right) \right]^{-1}
\label{eq;sampleTransmissionIndex}
\end{eqnarray}
where $d$ is the thickness of the film, $k=2 \pi n_{f}/\lambda$\, is the wavenumber within the film, and $n_{f}$ and $n_{s}$ are the indices of refraction of the film and substrate. There is one such equation for each circular polarization channel \textit{n} and \textit{p}.

Analysis of the data ultimately requires an equation for the index of refraction (or some other material property) in terms of the Faraday angle.   Equation~(\ref{eq;sampleTransmissionCircular}) with Eq.~(\ref{eq;sampleTransmissionIndex}) contain the pertinent information but the combination does not lend itself to inversion.  Fortunately, because the relative difference between the indices of refraction of the film for \textit{n} and \textit{p} polarizations is usually very small, we can generate a readily invertible form by expanding their combination about either $n_{f,n}$ or $n_{f,p}$.  Choosing $n_{f,p}=n_{f}$ with $\delta=n_{f,n}-n_{f,p}$, expanding Eq,~(\ref{eq;sampleTransmissionCircular}), and keeping only the linear term results in
\begin{equation}
\theta _{\text{F}}=\frac{-i\delta }{2}\frac{1}{t(n_{f}) }\frac{d\,t(n)}{dn}.
\label{eq;thetaFTransmissionApprox}
\end{equation}
Combining Eqs.~(\ref{eq;sampleTransmissionIndex}) and~(\ref{eq;thetaFTransmissionApprox}) produces
\begin{eqnarray}
\delta =2\theta_{\text{F}}\nonumber\ \ \ \ \ \ \ \ \ \ \ \ \ \ \ \ \ \ \ \ \ \ \ \ \ \ \ \ \ \ \ \ \ \ \ \ \ \ \ \ \ \ \ \ \ \ \ \ \ \ \ \ \ \ \ \ \ \ \ \\
\times \frac{n_{f}^2\left(n_{s}+1\right) -in_f\tan \left( kd\right) \left( n_{s}+n_{f}^{2}\right) }{kd\left( n_{s}+n_{f}^{2}\right)-\tan \left( kd\right) \left( n_{s}+ikd\left( n_{s}+1\right) n_{f}-n_{f}^{2}\right) }.\nonumber\\
\label{eq;deltanVersusThetaHApprox}
\end{eqnarray}
For an expression for $\sigma_{xx}$ and $\sigma_{xy}$ note that by the same symmetry assumed earlier the film dielectric permitivity tensor and the conductivity tensor are also diagonal in circular polarization, which we can convert to linear polarization as
\begin{eqnarray*}
&&\left\langle C\left|\sigma \right| C\right\rangle =\left( \begin{array}{cc}
\sigma _{p} & 0\\
0 & \sigma _{n}
\end{array}\right)\nonumber\\
&& =\left\langle C\left| L\right.\right\rangle \left\langle L\left| \left( \begin{array}{cc}
\frac{\sigma _{n}+\sigma _{p}}{2} & \frac{i\left( \sigma _{n}-\sigma _{p}\right) }{2}\\
\frac{i\left( \sigma _{p}-\sigma _{n}\right) }{2} & \frac{\sigma _{n}+\sigma _{p}}{2}
\end{array}\right) \right| L\right\rangle \left\langle L\left| C\right. \right\rangle\\
\end{eqnarray*}
Using Maxwell's equations the complex conductivity in CGS units is therefore
\begin{equation}
\sigma_{xx}=-\frac{i\omega}{8\pi}\left(n_{f,n}^2+n_{f,p}^2\right)=-\frac{i\omega}{4\pi}n_{f}^2
\label{eq;sigmxxVersusn}
\end{equation}
\begin{equation}
\sigma_{xx}=-\frac{i\omega}{8\pi}\left(n_{f,n}^2+n_{f,p}^2\right)=\delta\frac{\omega}{4\pi}
n_{f}
\label{eq;sigmxyVersusn}.
\end{equation}
By means of Eqs.~(\ref{eq;sigmxxVersusn}) and~(\ref{eq;sigmxyVersusn}), Eq.~(\ref{eq;deltanVersusThetaHApprox}) becomes
\begin{eqnarray}
\sigma_{xy}=\theta_{\text{F}} \frac{\omega n_{f}^2}{2 \pi }\ \ \ \ \ \ \ \ \ \ \ \ \ \ \ \ \ \ \ \ \ \ \ \ \ \ \ \ \ \ \ \ \ \ \ \ \ \ \ \ \ \ \ \ \ \ \ \nonumber&&\\
\times\frac{n_{f} \left (n_{s} +1 \right )-i\tan (kd) \left( n_{s} + n_{f}^2 \right) }{kd \left( n_{s}+n_{f}^2 \right)-\tan(kd)\left(n_{s}+ikd\left(1+n_{s}\right)n_{f}-n_{f}^2\right)}.\nonumber&&\\
&&\label{eq;sigmxyVesusthetaH}
\end{eqnarray}
This expression of course requires accurate values for the real and imaginary parts of $\theta_{\text{F}}$.  The experimental apparatus, however, introduces several errors into these values, which we have ignored up to this point but, which nevertheless must be removed or compensated as described below.

\subsection{Calibration}

There are three rather serious sources of error ignored in the foregoing.  First, the MCT detector and subsequent electronics introduce their own response functions which must be removed from the data.  Second, the PEM modulation reported by the PEM electronics is frequently in error for reasons to be discussed below.  Third, the Faraday angle $\theta_{\text{F}}$ determined by the above equations is the total Faraday angle, which includes a background, which results mostly from the ZnSe magnet windows and the substrate.

Regarding the response functions, Eqs.~(\ref{eq;thetafVersusPower}) and (\ref{eq;thetadVersusPower}) above require the ratio of the optical power in the 2nd and 3rd harmonics to the optical power at 112 Hertz (referred to as DC) at the MCT detector.  However, the MCT and other electronics possess frequency transfer functions, which attenuate these signals by different amounts, and once determined must be divided out of the collected data.  These functions were determined using a high-speed communications type laser diode to generate a known optical beam at the same frequency and amplitude as that within the experimental system.  When this beam was directed at the MCT the result corresponded rather well to the response of a simple RC circuit.

Next consider that the first lockin measures the RMS voltage of only the 1st harmonic of the 112 Hz square wave chopped signal, whereas Eqs.~(\ref{eq;thetafVersusPower}) and (\ref{eq;thetadVersusPower}) require the dc power. To determine this correction factor we examine the relation between a square wave and its first harmonic.  The normalization requirement generates the inverse of the first harmonic:
\begin{equation}
\int _{0}^{\tau }f^{-1}(t)\, f(t)\, dt=1
\end{equation}
Therefore, representing the harmonic as $\sin(\omega t)$, the normalization constant N is determined by:
\begin{equation}
\int _{0}^{\tau }f^{-1}(t)\, \sin (\omega t)\, dt=N\int _{0}^{\tau }f^{-1}(t)\, \sin (\omega t)\, dt=1
\end{equation}
Therefore $N=\frac{2}{\tau}$  and $f^{-1}(t)=\frac{2}{\tau}\sin (\omega t)$.  We now perform the integral transform to determine the first harmonic amplitude for a square wave of peak amplitude 1.
\begin{equation}
\int _{0}^{\tau }f^{-1}(t)\, (\text{square wave})\, dt=\frac{4}{\pi}
\end{equation}
So, for a square wave of peak amplitude 1, the output of the 7265 would be the RMS value for a sinewave of peak amplitude $4/\pi$.  Because data2$\omega$ and data3$\omega$ (also recorded in RMS) were divided by the 112 Hz RMS signal, removing this effect therefore entails multiplying them by $4/\pi$.

Next, consider the PEM.  The PEM electronics as supplied by the manufacturer at best only infer the amplitude of the phase modulation impressed upon the transmitted beam.  This inference does not consider the path of the beam through the ZnSe interaction crystal, the ZnSe crystal temperature, nor the possible changes in coupling between the driving quartz crystals and the ZnSe interaction crystal.  Equations~(\ref{eq;thetafVersusPower}) and (\ref{eq;thetadVersusPower}) , however, require an accurate knowledge of this modulation in the value of the variable $\beta$.  The current experiment explores, among other things, the frequency dependence of $\theta_{\text{F}}$.  Because the frequency dependence is often weak, its determination requires data with very high relative accuracy.  A simple method arises for obtaining $\text{Re}\left(\theta_{\text{F}} \right)$ which makes use of the provision for physically rotating the PEM assembly back and forth a predetermined amount as discussed above.  Prior to collecting each data set for each laser line one rotates the PEM assembly back and forth by the predetermined amount and records the readings from the lockin amplifiers.  This defines the calibration factor for $\text{Re}\left(\theta_{\text{F}} \right)$:
\begin{equation}
\text{C}_{\text{real}}=\frac{\text{predetermined\,physical\,rotation}}{\text{reading}_2-\text{reading}_1}
\end{equation}
where
\begin{equation}
\text{reading}_i=\frac{\text{reading}(2\omega)_i}{\text{reading}(112)_i}
\end{equation}
This calibration procedure also supplies the sign of $\text{Re}\left(\theta_{\text{F}}\right)$, which the electronics by itself leaves somewhat ambiguous.  Subsequent to this calibration the real part of the Faraday angle emerges from the data as
\begin{equation}
\text{Re}\left(\theta_{\text{F}}\right)=(\text{sign})\times \text{data}2\omega\times \text{C}_{\text{real}}.
\end{equation}

Because the frequency dependence of $\text{Im}\left(\theta_{\text{F}}\right)$ is also weak, its determination also requires data with very high relative accuracy.   Equation~(\ref{eq;thetafVersusPower}) along with the rotation calibration above also afford a relative calibration for $\theta_{\text{d}}$.  This calibration is essential in order to reliably determine the correspondingly small wavelength dependence of $\text{Im}\left(\theta_{\text{F}}\right)$ The procedure entails simply comparing the actual, predetermined, physical rotation with the calculated value from Eq.~(\ref{eq;thetafVersusPower}) and then adjusting the value of $\beta$, the retardance until they agree.  When calculating $\text{Im}\left(\theta_{\text{F}}\right)$ with Eq.~(\ref{eq;thetadVersusPower}) we use this adjusted value $\beta_{\text{adj}}$ in place of that determined from the manufacturers calibration.  This procedure should be performed initially, and then after each laser line change.  Here again the electronics leave the sign ambiguous, so it must be determined using a sample of known ellipticity such as a quarter waveplate.  Combining the foregoing calibration corrections:
\begin{equation}
\text{Im}\left(\theta_{\text{F}}\right)=(\text{sign})\frac{1}{\pi J_3\left(\beta_{\text{adj}}\right)}\text{data}3\omega(\lambda).
\end{equation}

Finally consider the background contribution to $\theta_{\text{F}}$ which manifests itself as a number of terms:
\begin{equation}
\theta_{\text{F,background}}=c_0+c_1B+c_2B^2+...
\end{equation}
where each coefficient $c_i$ may have a wavelength dependence.  In the current work the data sets contain $\theta_{\text{F}}$ as a function of magnetic field which is, for example, scanned from +8 to -8 Tesla or vice versa.  From this data we compute the slope $\partial \theta _{\text{F}}/\partial B$ which obviously does not contain $c_0$.  Further, because the magnetic field values are both positive and negative, computation of the average slope eliminates all terms even in {\bf B}.  Removing the remaining odd terms requires direct measurement of the background using a sample of the substrate material followed by simple subtraction of the result from the data sets.  The results of these measurements for substrates of Baf and high purity Si were purely real Faraday angles, which were consistent with the bandgap of these substrate materials and the ZnSe magnet housing windows as well as the absence of free carriers.

\section{Experimental Measurements}
\subsection{Semiconductors}

Faraday measurements were first performed on semiconductor samples for three reasons.
First, the signals are large and can be readily increased by using thicker samples.
Second, the optical properties of semiconductors are well known, which allows the accuracy
of the Faraday measurements to be verified by by comparing them to the results in other experiments.  Finally, since the 
thin film samples are grown on semiconductors, it is 
important to measure their contribution to the Faraday signals accurately in order to remove this 
background from the desired thin-film signal.   

The Faraday measurements were performed on three semiconductors. In semiconductors, two sources contribute to the Faraday signals \cite{Mavroides}.  The first source is from free carriers, which have a contribution to the Faraday signals that is proportional to $\omega^{-2}$, where $\omega$ is the radiation frequency.  The second source is from interband transitions, where the magnetic field  causes anisotropy in the refractive indices for left and right circularly polarized light.  This difference in index leads to a phase shift between left and right circularly polarized light, which in turn results in a rotation of the incident polarization (Faraday rotation).  For below band gap radiation, the interband contribution is proportional to $\omega^2$.  Since the MIR frequencies are an order of magnitude smaller than the energy band gap in typical semiconductors, no interband absorption occurs and therefore only the interband term contributes to Faraday rotation.  Furthermore, since the semiconductors in this experiment are of high purity and the MIR frequencies are relatively high, the free carrier contribution is minimized.  Faraday rotation caused by such samples is quantified in terms of its Verdet constant $V$, which is defined as the angle of rotation per unit magnetic field  per unit thickness of sample. The frequency dependence of the Verdet constant and its relationship to the Faraday angle at 8~T for a semiconductor is given by the following equation \cite{ruymbeek}:
\begin{equation}
V=u\omega^2+{v\over{\omega^2}}=\frac{\text{Re}\left[\theta_{\text{F}}(8~\text{T})\right]}{8~\text{T}\times D}\times\frac{180}{\pi}
\label{eq;verdet}
\end{equation}

\noindent where $u$ is the coefficient for the interband contribution and $v$ is the coefficient for the free carrier 
contribution to the Faraday rotation, and
$D$ is the sample thickness.
Ref.~\onlinecite{ruymbeek} \textit{et al.} have measured the Verdet constant as a function of
frequency (2000-3300 cm$^{-1}$ for several Si samples with different free carrier concentrations.
One can extend these results to 949 cm$^{-1}$ by using Eq.~\ref{eq;verdet} to obtain $u$ and $v$ from the data 
in Ref.~\onlinecite{ruymbeek}.  The Verdet coefficient at 949 cm$^{-1}$
is estimated to be $25.0~$degree/T$\cdot$m.  The value obtained at that frequency
by MIR Faraday measurements on a high purity ($>1$~k$\Omega$cm) silicon sample is $25.6$~degree/T$\cdot$m.   
This corresponds to 
Re$[\theta_{\text{F}}]=2\times 10^{-3}$ at 8~T and 949 cm$^{-1}$ for a 0.50~mm thick sample

\begin{figure*}
\includegraphics[clip=true]{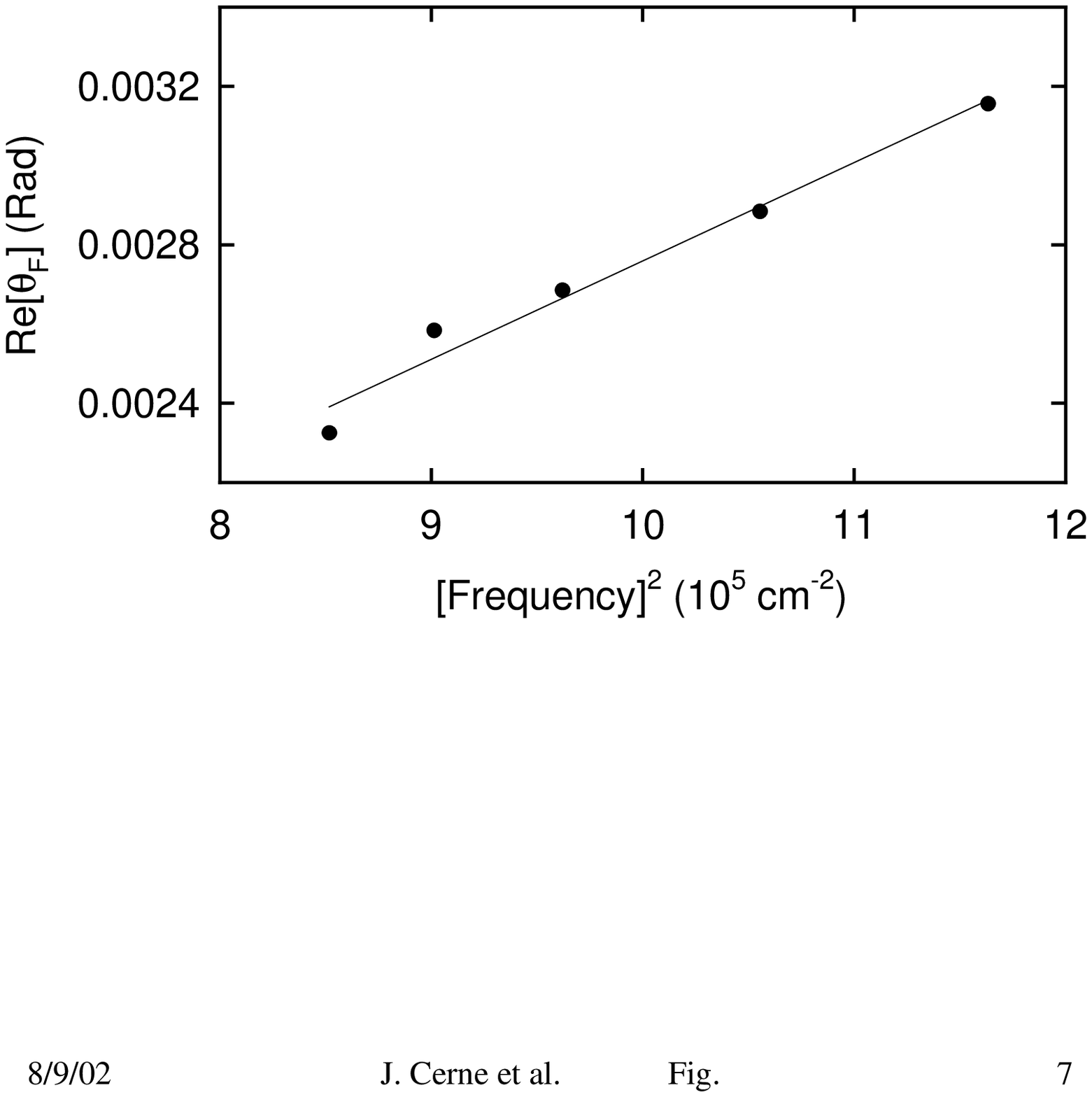}
\caption{\label{fig;gaas} The Re[$\theta_f$] as a function of frequency squared at 8~T for a GaAs sample.  The solid line represents a $\omega^2$-fit.  The semi-insulating sample is wedged with an average thickness of 0.42~mm.}
\end{figure*}

The second semiconductor material is semi-insulating GaAs.  In this case, the frequency dependence of the Faraday 
rotation signal is explored.  Fig.~\ref{fig;gaas} plots
the Re[$\theta_{\text{F}}$] as a function of frequency squared at 8~T for 
a GaAs sample.  The solid line represents a $\omega^2$-fit.  The semi-insulating 
sample is wedged with an average thickness of 0.42~mm.   The $\omega^2$ frequency behavior of Re[$\theta_{\text{F}}$] is 
consistent with Eq.~\ref{eq;verdet}.  At 8~T and 949 cm$^{-1}$,
the Re[$\theta_{\text{F}}$] is approximately $3\times10^{-3}$.  The Verdet coefficient is
approximately 44~degree/T$\cdot$m.  The value obtained at 10.59 $\mu$m for Cr-doped\cite{doped}
GaAs by Ref.~\onlinecite{phipps} is 41~degree/T$\cdot$m.
This is approximately 50~\% larger than that for Si, which is consistent with the fact that GaAs has direct band gap that 
is approximately a factor of two smaller than that of Si.
This brings the MIR radiation in GaAs closer to the refractive index anisotropy at the band edge.  In fact, the Re[$\theta_{\text{F}}$] for 
GaAs decreased as temperature is lowered, which is consistent with the increase in the band gap as temperature decreases.  As expected, 
no circular dichroism
signal is observed (Im$[\theta_{\text{F}}]\approx0$).

The final semiconductor investigated is LaSrGaO$_4$.  This material has the largest band
gap, and hence the smallest interband contribution to Re[$\theta_{\text{F}}$].  At 8~T and 949 cm$^{-1}$,
the Re[$\theta_{\text{F}}$] is approximately $6.5\times10^{-5}$ for a 0.31~mm thick sample.
This translates in a Verdet coefficient at 949 cm$^{-1}$ of 1.5~degree/T$\cdot$m.
Despite the onset of strong 
absorption by phonons for radiation below 1000 cm$^{-1}$, LaSrGaO$_4$ proved to be an excellent
substrate due to its weak contribution to the Re[$\theta_{\text{F}}$] background.
Note that as frequency is decreased, the strength of the semiconductor Faraday signals decreases (Eq.~\ref{eq;verdet}) 
while the signal from the free carriers in the thin-film metallic samples increases
(Eq.~\ref{eq;verdet}).  As a result, the lower frequency measurements are simpler and
more accurate.

\subsection{High Temperature Superconductor Bi$_2$Sr$_2$Ca$_1$Cu$_2$O$_{8+\delta}$}

\begin{figure*}
\includegraphics{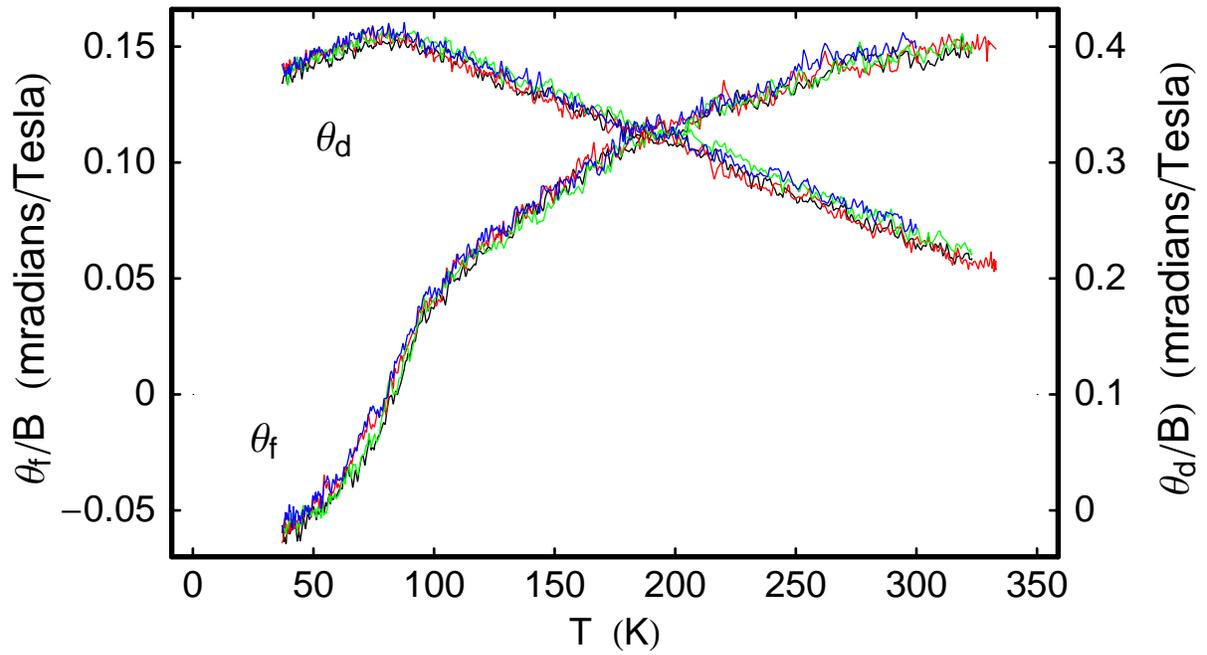}
\caption{\label{fig;BSCCOthetaFVersusT} The real and imaginary parts of the Faraday angle $\theta_f$ and $\theta_d$ each per Tesla and versus temperature for 2212 BSCCO measured at 950 cm$^{-1}$.  Each color represents a different subtracted pair of temperature scans.  There are four  pairs total in each graph.}
\end{figure*}

\begin{figure*}
\includegraphics{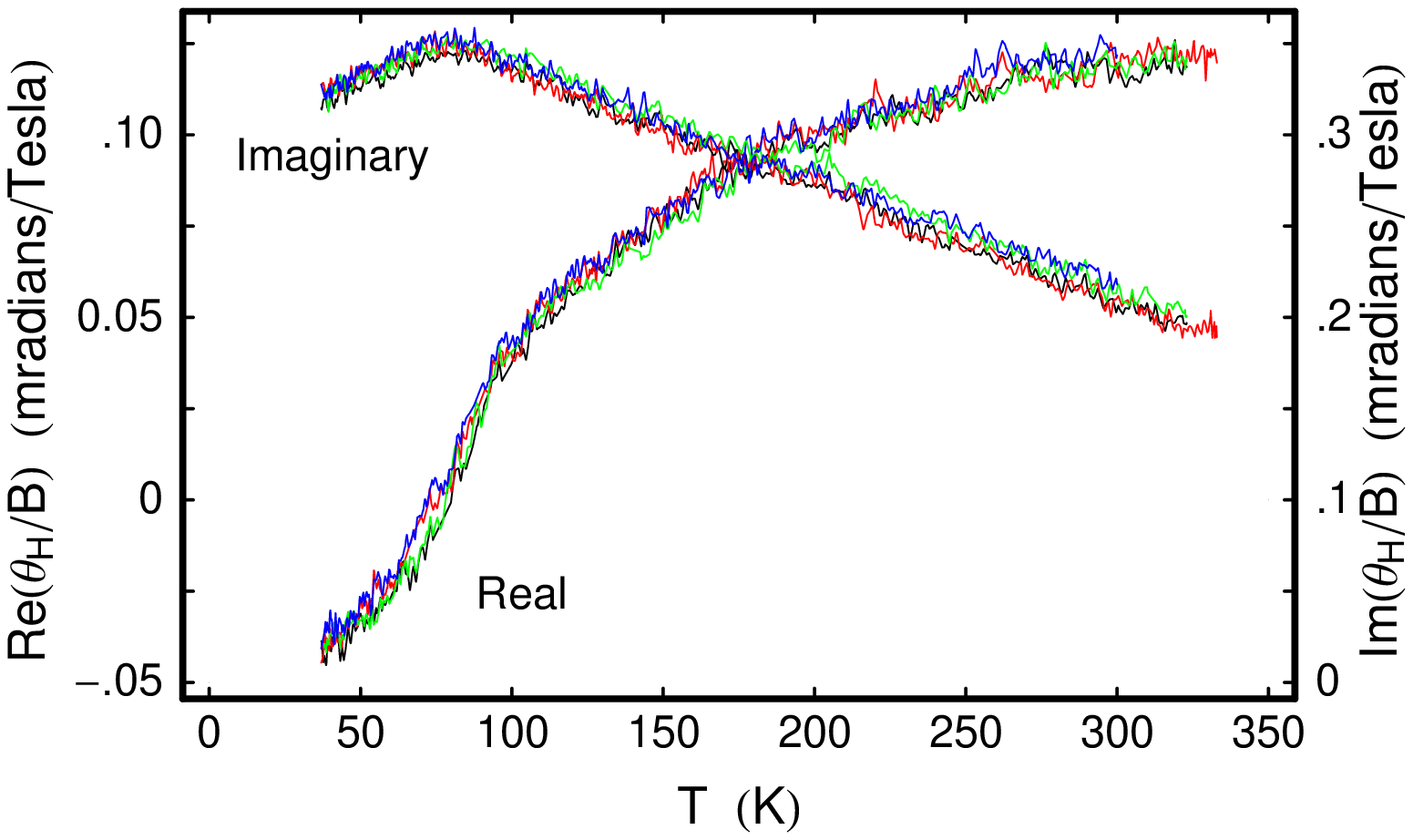}
\caption{\label{fig;BSCCOthetaHVersusT} The real and imaginary parts of $theta_H=\sigma_{xy}/\sigma_{xx}$ for 2212 BSCCO.  Each color represents a different subtracted pair of temperature scans all corresponding to a laser frequency of 950 cm$^{-1}$.  There are four  pairs in each graph.}
\end{figure*}

The measurements also include the response of Bi$_2$Sr$_2$Ca$_1$Cu$_2$O$_{8+\delta}$ to input radiation from 920 to 1090 cm$^{-1}$ and over a temperature range from 30 to 330 K in an external B field ranging from -8 to +8 Tesla.  Figure~\ref{fig;BSCCOthetaFVersusT} depicts result $\theta_{\text{F}}$ versus temperature.  This response is not too unlike the Hall angle $\theta_{\text{H}}=\sigma_{xy}/\sigma_{xx}$ with which it may be compared (see for example Fig.~\ref{fig;BSCCOthetaHVersusT}). To form $\theta_{\text{H}}$ one would usually divide $\sigma_{xy}$ by the value of $\sigma_{xx}$  corresponding to the same temperature and frequency.  But here, we wish to examine that part of the Hall angle related to the "free carriers."  Interband transitions which are on the order of  ~1 volt contribute essentially nothing to $\sigma_{xy}$.  However, we must remove their somewhat more substantial contribution to $\sigma_{xx}$ before taking the ratio.  In CGS units
\begin{equation}
\epsilon_{\text{total}}=\epsilon_{\text{bound}} + i \frac{4\pi\sigma_{\text{free}}}{\omega}=i\frac{4\pi\sigma_{\text{total}}}{\omega}
\end{equation}
\begin{equation}
\sigma_{\text{free}}=\sigma_{\text{total}}+\frac{i\omega\epsilon_{\text{bound}}}{4\pi}
\end{equation}
Figure~\ref{fig;BSCCOthetaHVersusT} displays the values of $\theta_{\text{H}}$ versus temperature.  The real part of $\theta_{\text{H}}$ increases with temperature but appears to saturate around 300 K.  Note that the real part of the Hall angle is positive for all temperatures in the normal state and does not show any striking discontinuity at T$_c$ at 90 K.  This data appears in Ref \cite{schmadel} where it is compared with far-infrared data from Ref \cite{Grayson}.  The results show a significant disconnect from the behavior of the Hall angle in the existing data for YBCO in the far-infrared, which indicate a negative value for the real part of the Hall angle above 250 cm$^{-1}$, whereas that of the current work at 1000 cm$^{-1}$ is positive.  The current work when analyzed using an extended Drude formalism results in a Hall mass comparable to the ARPES Fermi mass and a scattering rate comparable to the DC longitudinal, DC Hall, and far-infrared Hall scattering rates, which, however, are only 1/4 of the ARPES values.

\subsection{Metal Films}

\begin{figure*}
\includegraphics[clip=true]{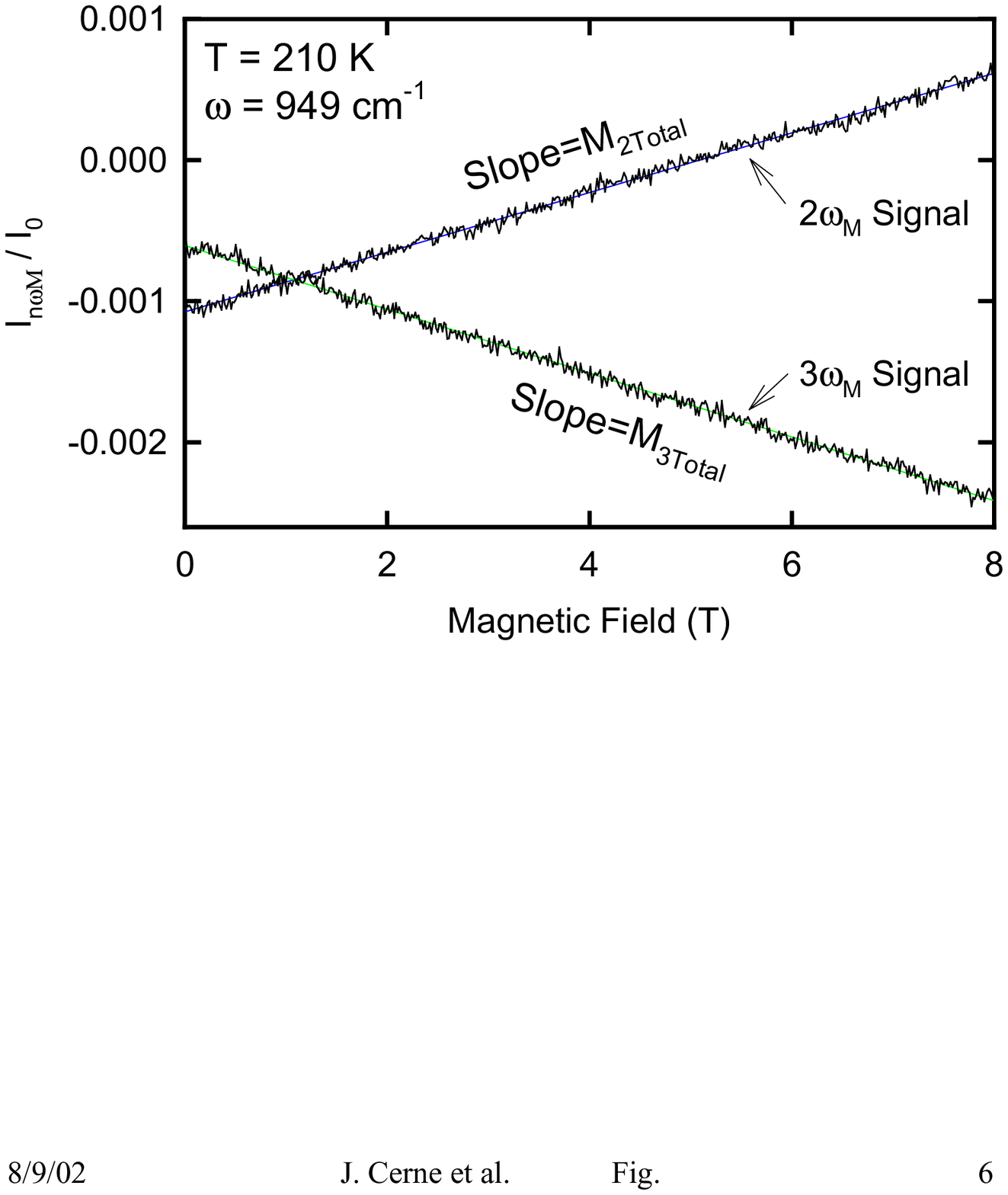}
\caption{\label{fig;far_signal} The $2\omega_{m}$ and $3\omega_{m}$ normalized signals for a Cu film as a function of magnetic field at 949 cm$^{-1}$ and 210~K.}
\end{figure*}

Figure~\ref{fig;far_signal} shows the $2\omega_{m}$ and $3\omega_{m}$ normalized
signals as a function of magnetic field
$B$ at room temperature for a Cu thin-film sample. The MIR radiation frequency is
949 cm$^{-1}$.  The  background contribution due to the substrate and windows has not been removed. Note the negative offset 
in the $2\omega_{m}$ data.   The signals are linear in $B$, as expected. 

It is interesting to compare the MIR Hall effect results on HTSC\cite{cerne,schmadel,rigal}
with those on Au and Cu films\cite{au,achall}. Surprisingly, the Hall angle results for both materials are qualitatively 
similar, and are well parameterized by the Drude form for $\theta_{\text{H}}$.  For both systems, the scattering rate $\gamma_{\text{H}}$ 
associated with the Hall angle is linear in temperature and independent of frequency.  In Au and Cu the longitudinal 
scattering rate $\gamma_xx$ (obtained from infrared conductivity measurements) is also temperature dependent and frequency
independent as expected from electron-phonon scattering when the
measurement frequency is higher than the Debye frequency. For optimally doped YBa$_2$Cu$_3$O$_7$ in the
MIR, however, while $\gamma_{\text{H}}$ has a strong temperature dependence and
no frequency dependence, $\gamma_xx$ is temperature independent but
frequency dependent (as found in infrared conductivity\cite{schlesinger} and angular-resolved photoemission 
spectroscopy\cite{arpes}). The behavior of
$\gamma_xx$ precludes phonons or magnons as the dominant scatterers and
the lack of frequency dependence of $\gamma_{\text{H}}$ is in contrast to the
predicted and observed behavior of a Fermi liquid or inelastic scattering
in general. Therefore the frequency and temperature dependence of
$\gamma_{\text{H}}$ that are reported in the MIR\cite{cerne} are highly unusual and
indicate a non-Fermi liquid behavior of the normal state of YBa$_2$Cu$_3$O$_7$.  Similar results have been obtained recently in other HTSC materials such as optimally doped Bi$_2$Sr$_2$CaCu$_2$O$_8$\cite{schmadel} and underdoped YBa$_2$Cu$_3$O$_7$.\cite{rigal}  
We note
that Ioffe and Millis \cite{ioffe} have recently proposed such a
relaxation rate behavior based on quasi-elastic scattering from
superconducting fluctuations in the normal state of high T$_c$ materials.
Fluctuation effects have also been observed in normal state of underdoped
Bi$_2$Sr$_2$CaCu$_2$O$_8$ in measurements of the longitudinal
conductivity by THz spectroscopy.\cite{orenstein}  The frequency and temperature dependence of the quasiparticle 
Hall response serves as a critical test for any theoretical model that attempts to explain the nature of the normal state in HTSC.

\begin{acknowledgments}

We  wish to estend our thanks to G. S. Jenkins, J. R. Simpson, and D. B. Romero for their assistance in performing the various reported measurements, and G. Gu and J. J. Tu  for supplying the BSCCO sample and transport data.

This work was supported by the NSF under contract no. DMR 0070959 

\end{acknowledgments}

\appendix

\section{Polarization Analysis Using Dirac Notation}

In polarimetric measurements a train of optical components, along with a sample, modify or operate on an input beam, which is in some state of known
intensity and polarization.  Optical components  typically include somewhat imperfect devices such as polarizers, waveplates, photoelastic modulators,
and mechanical rotators.  The modified output beam generally strikes a detector producing an output electrical signal.  Here we develop a general
formalism useful in relating the electrical signal to the properties of the sample.  The formalism uses Dirac notation which well differentiates
between a vector or state $|\psi\rangle$ and its representation in some basis: $\langle x|\psi\rangle$.  In polarimetry the usual bases are linear polarization and circular polarization.  The formalism also replaces the Jones
matrices with generalized operators and introduces the basis transformations and geometrical rotation operations.  Finally the result is related
to the output of the typical optical, ``square law'' detector.

In what follows the direction of propagation is in the positive $z$ direction and all angles are measured in the positive radial direction off the
$x$ axis and about the $z$ axis.  Beginning with the input beam, the formalism simply considers it to be a non-normalized ket
\begin{eqnarray*}
|\mathrm{in}\rangle.
\end{eqnarray*}
Expressed in the linear polarization basis it is
\begin{eqnarray*}
|\mathrm{in}\rangle=\sum_L|L\rangle\langle L|\mathrm{in}\rangle=
|x\rangle\langle x|\mathrm{in}\rangle + |y\rangle\langle y|\mathrm{in}\rangle
\end{eqnarray*}
or in the circular basis
\begin{eqnarray*}
|\mathrm{in}\rangle=\sum_C|C\rangle\langle C|\mathrm{in}\rangle=
|p\rangle\langle p|\mathrm{in}\rangle + |n\rangle\langle n|\mathrm{in}\rangle
\end{eqnarray*}
where $|p\rangle$ indicates an electric vector rotating in the positive radial direction about the $z$ axis as seen at a fixed point and $|n\rangle$ indicates an electric vector rotating in the negative radial direction about the z axis also as seen at a fixed point. $|p\rangle$ and $|n\rangle$ are also referred to as having positive and negative helicity.  In matrix notation
\begin{eqnarray*}
\langle L|\mathrm{in}\rangle=
\pmatrix{
E_x\cr
E_y\cr
}
\qquad
\mathrm{and}
\qquad
\langle C|\mathrm{in}\rangle=
\pmatrix{
E_p\cr
E_n\cr
}.
\end{eqnarray*}
The various components in the optical train appear as operators acting on the input ket to produce the output ket:
\begin{eqnarray*}
|\mathrm{out}\rangle={\bf TVW}|\mathrm{in}\rangle.
\end{eqnarray*}
In the linear basis this is
\begin{eqnarray*}
|{out}\rangle\,=&&|L\rangle\langle L|TVW|L'|\mathrm{in}\rangle\\=&&|L\rangle\langle L|T|L''\rangle\langle L''|V|L'''\rangle\langle L'''|W|L'\rangle\langle L'|\mathrm{in}\rangle\\=&& |L\rangle\langle L|T|L\rangle\langle L|V|L\rangle\langle L|W|L\rangle\langle L|\mathrm{in}\rangle
\end{eqnarray*}
where summation is assumed over repeated state designations L; and the primes have been dropped with the understanding the order of the matrices will not be changed.

Some devices or operations such as geometric rotation and Faraday rotation are most easily represented in the circular basis.    The transformation
matrices, $\langle C|L\rangle$ and $\langle L|C\rangle$ , are
the means of conversion between these two bases.  The transformation for states proceeds as
\begin{eqnarray*}
|\mathrm{in}\rangle =&& \sum_L|L\rangle\langle L|\mathrm{in}\rangle\\
=&& \sum_{LC}|C\rangle\langle C|L\rangle\langle L|\mathrm{in}\rangle\\
=&&|C\rangle\langle C|L\rangle\langle L|\mathrm{in}\rangle ,
\end{eqnarray*}
and for operators as
\begin{eqnarray*}
{\bf T} =&& \sum_{L,L'}|L\rangle\langle L|{\bf T}|L'\rangle\langle L'|\\
=&& \sum_{L,L'}\sum_{C,C'}|C\rangle\langle C|L\rangle\langle L|{\bf T}|L'\rangle\langle L'|C\rangle\langle C'|\\
=&& |C\rangle\langle C|L\rangle\langle L|{\bf T}|L\rangle\langle L|C\rangle\langle C|
\end{eqnarray*}
In matrix representation
\begin{eqnarray*}
\langle C|L\rangle={1\over{\sqrt2}}
\pmatrix{
1 & -i \cr
1 & i \cr
}
\ \ 
\mathrm{and}
\ \ 
\langle L|C\rangle={1\over{\sqrt2}}
\pmatrix{
1 & 1 \cr
i & -i \cr
}.
\end{eqnarray*}
Below are the representations of some simple optical components in the linear bases:
a perfect polarizer aligned along the x axis
\begin{eqnarray*}
\pmatrix{
1 & 0 \cr
0 & 0 \cr
},
\end{eqnarray*}
a 1/4 waveplate aligned with slow axis along x axis
\begin{eqnarray*}
\pmatrix{
e^{i{\pi\over2}} & 0 \cr
0 & 1 \cr
},
\end{eqnarray*}
a photoelastic modulator aligned along the x axis
\begin{eqnarray*}
\pmatrix{
e^{i\beta\cos(2\pi ft)} & 0 \cr
0 & 1 \cr
}.
\end{eqnarray*}

Another concern for the devices or components is their orientation.  Assume, for example, that the operator $T$ represents the effect of some component
such a polarizer which is aligned with its critical direction at an angle $\theta$ from the $x$ axis.  A simple rotation operation develops an expression for the rotated polarizer:
\begin{eqnarray*}
T(\theta)={\bf RTR}^{-1}
\end{eqnarray*}
where $R$ is an active rotation and in circular representation is
\begin{eqnarray*}
\langle C|{\bf R}(\theta)|C\rangle=\pmatrix{
e^{-i\theta} & 0 \cr
0 & e^{i\theta} \cr
}.
\end{eqnarray*}

Finally, the output of a square-law type optical detector, aside from any responsivity factor is for our example
\begin{eqnarray*}
d(t)\propto\langle\mathrm{out}|\mathrm{out}\rangle
\end{eqnarray*}
where $|out\rangle$ is the Hermitian conjugate of $\langle out|$.

\section{Multilayer Response Using Relative Impedance Matrices}

Assume that some layered materials are arranged normal to the z axis and we desire a formalism for the transmitted and reflected intensities for
plane waves incident at arbitrary angles.  To begin, the plane wave solution to Maxwell's equations is:
\begin{eqnarray}
\vec{E}_{i,\pm} =&& \vec{\mathcal{E}}_{i,\pm} e^{i\vec k_{i,\pm}\cdot\vec r-i\omega t}
\label{eq;EMaxSol}\\
\vec{H}_{i,\pm} =&& \vec{\mathcal{H}}_{i,\pm} e^{i\vec k_{i,\pm}\cdot\vec r-i\omega t}
\label{eq;HMaxSol}
\end{eqnarray}
where   $+$ represents a wave traveling in the positive $z$ direction at some otherwise arbitrary angle, $-$ represents a wave traveling in the negative $z$ direction, and $i$ is the layer number. Then in CGS:
\begin{eqnarray}
\vec H_{i,\pm}=&& \sqrt{\epsilon_i\over{\mu_i}}\quad\hat k_{i,\pm}\times\vec E_{i,\pm}\nonumber\\
=&&{1\over{Z_i}}\quad\hat k_{i,\pm}\times\vec E_{i,\pm}\qquad\qquad\mathrm{CGS}
\label{eq;HVersusE}
\end{eqnarray}
where $\hat k_{i,\pm}$ is a unit vector parallel to the propagation direction, $\epsilon_i$ is the complex dielectric constant of the $i$th layer and includes all conductivity effects and $Z_i=\sqrt{\epsilon_i\over{\mu_i}}$ is the relative impedance of the medium and like $\epsilon$ may be complex.

Now that we have a solution inside a layer we need to match solutions for different layers at the boundaries between layers.  The curl $E$ Maxwell equation provides one boundary condition:
\begin{equation}
\vec E_i\times\hat z=\vec E_{i+1}\times\hat z.
\label{eq;EiVersusEi+1}
\end{equation}
The curl $H$ Maxwell equation provides the other boundary condition:
\begin{equation}
\vec H_i\times\hat z=\vec H_{i+1}\times\hat z.
\label{eq;HiVersusHi+1}
\end{equation}

To apply the foregoing consider first S polarization ($E$ perpendicular to the plane of incidence) at each side of a boundary.  In such an instance
the $E$ field is tangent to the boundary and if we orient the $x$ and $y$ axes so that $E$ is also perpendicular to the $y$ axis, then, from Eq.~\ref{eq;EiVersusEi+1}:
\begin{equation}
\vec{\mathcal{E}}_{i,+,x}+\vec{\mathcal{E}}_{i,-,x}=
\vec{\mathcal{E}}_{i+1,+,x}+\vec{\mathcal{E}}_{i+1,-,x}.
\label{eq;Ei+Ei-VersusEi+1+Ei+1-}
\end{equation}
Likewise, substituting Eq.~\ref{eq;HVersusE} into Eq.~\ref{eq;HiVersusHi+1} produces:
\begin{equation}
{\cos(\theta_i)\over{Z_i}}(\vec{\mathcal{E}}_{i,+,x}-\vec{\mathcal{E}}_{i,-,x})
={\cos(\theta_{i+1})\over{Z_{i+1}}}(\vec{\mathcal{E}}_{i+1,+,x}-\vec{\mathcal{E}}_{i+1,-,x})
\label{eq;EithetaVersusEi+1theta}
\end{equation}
where $\theta_i$ is the angle of incidence measured of the z axis. Defining the incident impedance for S polarization as:
\begin{equation}
Z_{S,i}={\cos(\theta_i)\over{Z_i}}
\label{eq;ZS}
\end{equation}
Eq.~\ref{eq;EithetaVersusEi+1theta} becomes
\begin{equation}
Z_{S,i}(\vec{\mathcal{E}}_{i,+,x}-\vec{\mathcal{E}}_{i,-,x})
=Z_{S,i+1}(\vec{\mathcal{E}}_{i+1,+,x}-\vec{\mathcal{E}}_{i+1,-,x}).
\label{eq;Ei+Ei-,ZSVersusEi+1+Ei+1-,ZS}
\end{equation}
Combining Eqs.~\ref{eq;Ei+Ei-VersusEi+1+Ei+1-} and ~\ref{eq;Ei+Ei-,ZSVersusEi+1+Ei+1-,ZS} into a matrix form:
\begin{eqnarray}
\pmatrix{
{\mathcal{E}}_{i+1,+} \cr
{\mathcal{E}}_{i+1,-}\cr
}
= {\bf S}_{i,i+1}
\pmatrix{
{\mathcal{E}}_{i,+} \cr
{\mathcal{E}}_{i,-}\cr
}.
\label{eq;EandS}
\end{eqnarray}
where
\begin{eqnarray}
{\bf S}_{i,i+1}
= {1\over{2Z_{s,i}}}
\pmatrix{
Z_{s,i}+Z_{s,i+1} & Z_{s,i}-Z_{s,+i} \cr
Z_{s,i}-Z_{s,i+1} & Z_{s,i}+Z_{s,+i} \cr
}
\label{eq;S}
\end{eqnarray}

For P polarization (E field $\|$ to plane of incidence) define the incident impedance as:
\begin{equation}
Z_{p,i}=\cos(\theta_i)Z_i
\label{eq;ZP}
\end{equation}
then following the same procedure:
\begin{eqnarray}
\pmatrix{
{\cal H}_{i+1,+} \cr
{\cal H}_{i+1,-}\cr
}
= P_{i,i+1}
\pmatrix{
{\mathcal{H}}_{i,+} \cr
{\mathcal{H}}_{i,-}\cr
}
\label{eq;HandP}
\end{eqnarray}
where
\begin{eqnarray}
P_{i,i+1} = {1\over{2Z_{p,i}}}
\pmatrix{
Z_{p,i+1}+Z_{p,i} & Z_{p,i+1}-Z_{p,i} \cr
Z_{p,i+1}-Z_{p,i} & Z_{p,i+1}+Z_{p,i} \cr
}
\label{eq;P}
\end{eqnarray}
The $E$ field can be determined from $H$ using Eq.~\ref{eq;HVersusE}.

Equations~\ref{eq;ZS},~\ref{eq;EandS}, and~\ref{eq;S} and Eqs.~\ref{eq;ZP},~\ref{eq;HandP} and~\ref{eq;P} provide the incident and reflected amplitudes in the $i+1$ layer at the interface
in terms of the amplitudes in the $i$ layer at the interface for S and P polarization respectively.   It remains to develop the equations
to propagate the amplitudes across layer $i$ of thickness $d_i$.  This is already provided by Eqs.~\ref{eq;EMaxSol} and~\ref{eq;HMaxSol} which in matrix
form become:
\begin{eqnarray}
\pmatrix{
{\mathcal{E}}_{i,+}(z=d_i) \cr
{\mathcal{E}}_{i,-}(z=d_i)\cr
}&=&\pmatrix{
e^{i\vec k_{i,\pm}\cdot\hat z} & 0 \cr
0 & e^{i\vec k_{i,\pm}\cdot\hat z} \cr
}
\pmatrix{
{\mathcal{E}}_{i,+}(z=0) \cr
{\mathcal{E}}_{i,-}(z=0) \cr
}\nonumber\\
&=&{\bf U}_i
\pmatrix{
{\mathcal{E}}_{i,+}(z=0) \cr
{\mathcal{E}}_{i,-}(z=0) \cr
}\nonumber
\end{eqnarray}

So, as an example, given a set of layers numbered 1 to n from left to right the equation relating the incident and reflected amplitudes on left side of the set to the right side for S polarization is:
\begin{eqnarray}
{\bf S}_{n-1,n}{\bf U}_{n-1}{\bf S}_{n-2,n-1}\ldots {\bf U}_{i+1}{\bf S}_{i,i+1}
\pmatrix{
{\mathcal{E}}_{i,+}(left) \cr
{\mathcal{E}}_{i,-}(left) \cr
}
=\nonumber\\
\pmatrix{
{\mathcal{E}}_{n,+}(right) \cr
{\mathcal{E}}_{n,-}(right) \cr
}\nonumber
\end{eqnarray}

\begin{thebibliography}{99}
\bibitem{doped} Since the free carrier contribution to $V$ falls of with frequency
as $\omega^{-2}$, we do not expect this doping to strongly affect $V$ at 1000 cm$_{-1}$.  Therefore,
measurements of $V$ on doped and undoped GaAs should be comparable at 1000 cm$_{-1}$.
\end{thebibliography}
\end{document}